\documentstyle[11pt,epsf,twoside,cite]{article}

\def\ltap{\raise.3ex\hbox{$<$\kern-.75em\lower1ex\hbox{$\sim$}}}
\def\gtap{\raise.3ex\hbox{$>$\kern-.75em\lower1ex\hbox{$\sim$}}}
\def\smt{SU(3) $\times$ SU(2) $\times$ U(1)}
\def\ie{{\it i.e.}}
\def\etal{{\it et al.}}
\def\eg{{\it e.g.}}
\def\cfr{{\it cfr.}}
\def\n#1{\tilde{N}_{#1}}
\def\c#1{\tilde{C}_{#1}}
\def\cp#1{\tilde{C}^+_{#1}}
\def\cm#1{\tilde{C}^-_{#1}}

\def\se#1{\tilde{e}_{#1}}

\def\sepm#1{\tilde{e}^{\pm}_{#1}}
\def\semp#1{\tilde{e}^{\mp}_{#1}}
\def\smu#1{\tilde{\mu}_{#1}}

\def\sle#1{\tilde{\ell}_{#1}}

\def\sta#1{\tilde{\tau}_{#1}}
\def\miss#1{\not\!\! #1}
\def\G{\tilde{G}}
\renewcommand{\to}{\rightarrow}

\def\NPB#1#2#3{Nucl. Phys. {\bf B#1} (19#2) #3}
\def\PLB#1#2#3{Phys. Lett. B {\bf #1} (19#2) #3}
\def\PLBold#1#2#3{Phys. Lett. {\bf #1B} (19#2) #3}
\def\PRD#1#2#3{Phys. Rev. D {\bf #1} (19#2) #3}
\def\PRDold#1#2#3{Phys. Rev. {\bf D#1} (19#2) #3}
\def\PRL#1#2#3{Phys. Rev. Lett. {\bf #1} (19#2) #3}

\begin{document}
\pagestyle{myheadings}
\markboth
{\it S.~Ambrosanio and A.~E.~Nelson / New Multi-Scale SUSY Models...}
{\it S.~Ambrosanio and A.~E.~Nelson / New Multi-Scale SUSY Models...}

\begin{titlepage}
\thispagestyle{empty}
\begin{flushright}
UW/PT-97/11 \\
hep-ph/9707242 \\
\end{flushright}

\vspace{2.0cm}

\begin{center}
{\Large\bf
New Multi-Scale Supersymmetric Models with \\
\vspace{0.2cm}
Flavor Changing Neutral Current Suppression \\
}
~\\
{\large  Sandro Ambrosanio$^a$ and  Ann E. Nelson$^b$} \\
~\\
\noindent
{\it\ignorespaces
       (a) Randall Physics Laboratory, University of Michigan,
           Ann Arbor, MI 48109-1120, USA\\
       (b) Department of Physics 1560, University of Washington,
           Seattle, WA 98195-1560, USA\\
}

\vspace*{\fill}

{\bf Abstract} \\
\end{center}
{\small
We discuss the phenomenology of a class of supersymmetric models in
which some of the quark and lepton superfields are an integral part of
a dynamical supersymmetry breaking sector. The corresponding squarks and
sleptons are much heavier than any other superpartners, and could
naturally have masses as high as $\sim 40$~TeV. We discuss a general set
of conditions for acceptable flavor-changing neutral currents and
natural electroweak symmetry breaking, and identify two particularly
interesting new classes of theories. We discuss how  phenomenological
signatures of such multi-scale models at the CERN LEP~II and Fermilab
Tevatron colliders could significantly differ from previously considered
scenarios. In particular, we give experimental signals which could be present
if the left-handed selectron is much lighter than the right-handed one.
}

\vspace*{\fill}

\noindent
\parbox{0.45\textwidth}{\hrule\hfill} \\
\begin{tabular}{cl} e-mail: &  {\sl ambros@umich.edu,} \\
                            &  {\sl anelson@phys.washington.edu}
\end{tabular}
\end{titlepage}

\setcounter{page}{0}
\thispagestyle{empty}
\null
\newpage

\thispagestyle{plain}
\null
\vspace{-1.0cm}
A compelling solution to the gauge hierarchy problem is that
the world is supersymmetric at short distances. Testing this
hypothesis directly requires a discovery of the superpartners,
which requires understanding their experimental signatures,
which in turn depend on the supersymmetric spectrum.
However we still lack a standard model of supersymmetry (SUSY)
breaking, and a general phenomenological approach for SUSY
breaking (SSB) introduces an extravagant number of free parameters.
Thus, unless one has unambiguous data for physics beyond the standard
model (SM), searches for SUSY inevitably involve numerous assumptions.

Even if all sparticles are out of near-term experimental reach,
supersymmetrizing the SM can lead to observable
effects such as lepton flavor violation (LFV), particle electric
dipole moments (EDMs), and flavor-changing neutral currents (FCNCs).
Many explanations for the non-observation of such effects appeared 
in the literature
\cite{oldGMSB,newGMSB,dine,dg,pom,CKN,dp,nw,nonabel,align}, and these
also have the advantage of reducing the number of parameters needed to
describe the superpartner masses and couplings. Clearly, given our lack of
understanding of SSB, all schemes which account for the absence of flavor
violation deserve serious study.

In general, the explanations which have been given previously fall into two
classes.

\begin{enumerate}
\item{\bf Approximate Global Symmetries }
can constrain the SSB parameters. Suppression of FCNCs and LFV
can  arise either as a result of spontaneously broken horizontal flavor
symmetries \cite{nonabel,align}, or simply as a consequence of
accidental approximate flavor symmetries of the SSB and mediation sectors
\cite{oldGMSB,newGMSB}.

\item{\bf Decoupling}
of the first two generations of superpartners can suppress EDMs, FCNCs
and LFV \cite{dine,dg,pom,CKN,dp,nw}. The sparticles which potentially
mediate unacceptable flavor violation are those of the first two quark
and lepton generations. The principle of naturalness (\ie, that
fine-tuned cancellations between different terms should only occur as
result of a symmetry) is usually used to argue that all superpartners
should be lighter than $\sim 1$~TeV in order to explain the electroweak (EW)
scale. However since the first two generations are only weakly coupled
to the Higgs, the first two generations of scalars can naturally be much
heavier without spoiling natural EW symmetry breaking (EWSB)%
\footnote{If the SSB scale is too high, large logarithms can spoil
          natural EWSB \cite{dg}, or give negative masses squared to
          the lighter squarks and sleptons \cite{logs}.}.
Naturalness still requires the top squarks, the left-handed (L) bottom
squark, the EW gauginos and the higgsinos to be lighter than $\sim 1$~TeV.
Such a hierarchy in the superpartner spectrum could be a result of new
gauge interactions which are carried by the first two generations and
which are involved in dynamical SSB \cite{CKN,dp}.
\end{enumerate}

More generally, a hybrid of the above two solutions is possible.
The FCNC constraints can be satisfied provided that,
for the first two generations, the  superfields with the same \smt\
gauge quantum numbers  either both strongly couple to SSB dynamics
(so the scalar superpartners are very heavy) or both couple to the
SSB sector weakly, but in an approximately flavor blind way.
If, \eg, the L down type squarks are light and degenerate,
and the right-handed (R) down type squarks are non-degenerate but very
heavy, superpartner contributions to $\Delta S=2$ and $\epsilon_K$ are
sufficiently suppressed.

Such a hybrid solution can be the result of dynamical SSB at relatively
low energies (below $\sim 10^{10}$~GeV) if some of the quarks and
leptons superfields take  part in the SSB dynamics. For example, in the
model of Ref.~\cite{strassler}, the hierarchical structure of quark and
lepton masses is explained by  making the ``10's'' (of SU(5) \cite{GG})
composite, while the ``$\bar 5$'s'' are elementary. If the compositeness
dynamics  is also responsible for dynamical SSB \cite{CKN,dsbcomp},
the composite squarks and sleptons will be much heavier than the other
sparticles \cite{CKN}. The fundamental superfields are only weakly
coupled to the SSB sector. Since SSB is most effectively communicated to
the fundamental superpartners via \smt\ gauge dynamics \cite{oldGMSB,newGMSB},
the light sparticles with the same gauge quantum numbers will be nearly
degenerate. More generally, such a spectrum could result in any model with
gauge-mediated SSB (GMSB) and new horizontal gauge interactions carried
by some quarks and leptons, as well as the SSB sector.

This hybrid scenario can be even more effective at suppressing
FCNCs in a natural way than the pure decoupling scenario, for two reasons.
First, the strongest bound on flavor and $CP$ violation in the SUSY
parameters comes from the contribution of the ``left-right'' (L-R)
operator $\bar d_L s_R \bar d_R s_L$ to $K \bar K$ mixing and
$\epsilon_K$, and this is suppressed as long as  either L or R down-type
squarks are degenerate.
Second, the naturalness upper bound of $\sim 20$~TeV on the superpartner
masses assumes that all superpartners of the first two generations are 
heavy. Actually, the naturalness bound, which comes from two loop graphs
involving SU(2) gauge fields, is on the average mass squared of SU(2) 
doublets. For example, the R $d$-squarks and L sleptons could naturally 
be as heavy as $\sim 40$~TeV, provided the other scalars are lighter.

Theories with non-universal scalar masses which are
larger than $\sim 1$~TeV potentially generate a disasterous large
Fayet-Iliopoulos (FI) term \cite{fi} for weak hypercharge \cite{dg}.
The simplest way to avoid a FI term is an SU(5) approximate global
symmetry of the SSB dynamics. Thus, we will assume that the heavy
sparticles come in complete SU(5) multiplets with nearly degenerate masses.

With this SU(5) assumption, there are two classes of SSB theories which
avoid FCNCs and FI terms, allow natural EWSB, and which have never been
previously discussed. In the following, we will refer to such a theory
as a Hybrid Multi-Scale Supersymmetric Model (HMSSM).

{\bf  HMSSM-I:} All the sparticles are lighter than $\sim 1$~TeV,
except for the scalar superpartners of the first two generations
which transform as 10's (L squarks, R up and
charm squarks, and R sleptons). The light sparticles with
identical \smt\  quantum numbers are assumed to be nearly degenerate.

{\bf  HMSSM-II:}  All the sparticles are lighter than $\sim 1$~TeV,
except for the   scalar superpartners which transform as a $\bar 5$'s
(R down squarks and L sleptons). We further
break this class down into
HMSSM-IIa,
  where the R bottom squark and L $\tau$
  and $\nu_\tau$ sleptons are heavy, and
HMSSM-IIb,
  where the R bottom squark and L $\tau$
  and $\nu_\tau$ sleptons are light.
The light sparticles of the first two generations with
identical \smt\ quantum numbers are nearly degenerate.

 The HMSSM-I and HMSSM-IIb  are distinguished from the HMSSM-IIa
by treating the third generation differently from the first two.
As a consequence, there are potentially detectable new contributions to
$CP$ violation in $B$ decays \cite{bphys} and lepton flavor violation,
while such effects will be much smaller in the HMSSM-IIa.

 In the following, we explore the effects of multiple scales
for  superpartner masses on SUSY searches at LEP~II and Tevatron.
A complete examination is beyond the scope of this
letter, so we will just give a few illustrative examples of some
possible dramatic effects, mainly within the HMSSM-I.
We will assume the lightest superpartner (LSP) is the
gravitino ($\G$). However the fundamental SSB scale may be
high enough ($\gtap\ 10^6$~GeV) so that
decays into the $\G$ do not occur inside the detector. Motivated by
our assumption that the SUSY breaking masses for gauginos and
first two generation light superpartners are mainly gauge mediated,
we assume the conventional grand-unified relations amongst
the gaugino masses and, for the light sparticles, a spectrum consistent
with a general GMSB pattern, \eg\ colored superpartners are substantially
heavier than color neutral ones.
The SUSY signatures depend dramatically on which sparticle(s)
can decay dominantly to the $\tilde{G}$ (see, \eg, Ref.~\cite{AKM97}).
Often, this is only the case for the next-to-lightest superpartner (NLSP).
With our assumptions, in the HMSSM-I, the NLSP is either a neutralino, or
the lightest (most likely R) tau slepton $\sta{1}$, or (one of) the sneutrinos.
More complex (``co-NLSP'') scenarios with more than one sparticle
decaying directly into the $\G$ might occur as well.
In the HMSSM-II, the (co-)NLSP can be a neutralino and/or the
(R) $\sta{1}$, or all the R sleptons could be co-NLSPs. Furthermore,
for the HMSSM-IIb only, the tau sneutrino can play a (co-)NLSP role.
In general, for $\tan\beta$ not too large ($\ltap$ 4
to 20, depending on the model details), the lightest neutralino
$\n{1}$ is usually the NLSP. For simplicity, we mainly
consider this possibility. However, in all models, the $\sta{1}$
is amongst the light sparticles, and with our assumptions, it is always
one of the lightest scalars. Therefore, in any version of the HMSSM,
a signal from $\sta{1}\sta{1}$ production is a prime candidate for
discovery at LEP~II. If $\sta{1}$ is the NLSP,
such a signal could feature energetic, central tau leptons and large
$\miss{E}$ or tracks from massive, long-lived charged particles,
depending on whether the SSB scale is low ($\ltap\ 10^5$~GeV) or much higher.
However, such signatures can also arise from conventional GMSB models
\cite{AKM97,DTW2,DDN}.

When $\n{1}$ is the NLSP, the model phenomenology and the existing
bounds from experimental data crucially depend on whether the decay
$\n{1}\to\G\gamma$ occurs (i) promptly, (ii) (mostly) inside the detector,
or (iii) outside, which in turn depends on the SSB scale. Case (i) is of
special interest, leading to a variety of unusual events with
$\gamma\gamma + X + \miss{E}$ at colliders
\cite{DTW2,DDN,eegg1,2gamma,AKM96,DTW1,BKW,BBCT}.
One of them, where $X =$ two charged leptons, might have
been observed at Fermilab by CDF \cite{eegg1,theevent,eegg2}.
In case (i) [(ii)], an inclusive search for events with two energetic,
central [displaced] photons and $\miss{E_T}$ in the present
Tevatron data sample \cite{CDF-D0} can {\it at least} exclude masses
for $\n{1}$ and the lightest chargino $\c{1} \; \ltap\ 70$ and 125~GeV,
respectively, in a model independent way \cite{AKM96}.
Thus, the only fermion sparticle production process within LEP~II reach
is NLSP pair production.
A further consequence of the $\n{1}$ and $\c{1}$ mass lower limits is that,
within the parameter space of interest for SUSY at LEP~II ($m_{\n{1}}\;
\ltap$ 95~GeV), they select a region where $\n{1} \sim$
B-ino ($\tilde{B}$), which couples most strongly to the $\se{R}$.
Also, the relations $m_{\c{1}} \sim m_{\n{2}} \sim 2 m_{\n{1}}$
turn out to be always fulfilled. All these considerations must
also apply to the HMSSM, since they only come from assumptions
about the NLSP and the SSB scale, in addition to gaugino mass
unification.

For example, consider the HMSSM-I where L slepton pairs can be produced 
in a future LEP~II run, \ie\ $m_{\se{L}}$, $m_{\smu{L}} < 95$~GeV, while
the R selectron $\se{R}$ is very heavy%
\footnote{Actually, none of the methods given in this paper can
          distinguish between a moderately heavy selectron,
          $m_{\se{}} \sim 300$~GeV, and the ultra heavy ($\gtap\; 5$~TeV)
          mass expected in a multi-scale model. Measurement of
          ``super-oblique'' parameters \cite{superoblique} could 
          eventually further constrain the heavy selectron mass.}.
We also assume that the $\n{1}$ is the NLSP,
and that the $\n{1}$ decays into a photon and gravitino within the
detector. Given a sufficient number of selectron events at LEP~II,
it is possible to use the total cross sections and forward-backward
(F-B) asymmetry to deduce that that the selectron events must be
left-handed and therefore that the R selectron must be heavier.
\begin{figure}[h]
\centerline{
\epsfxsize=\textwidth
\epsffile{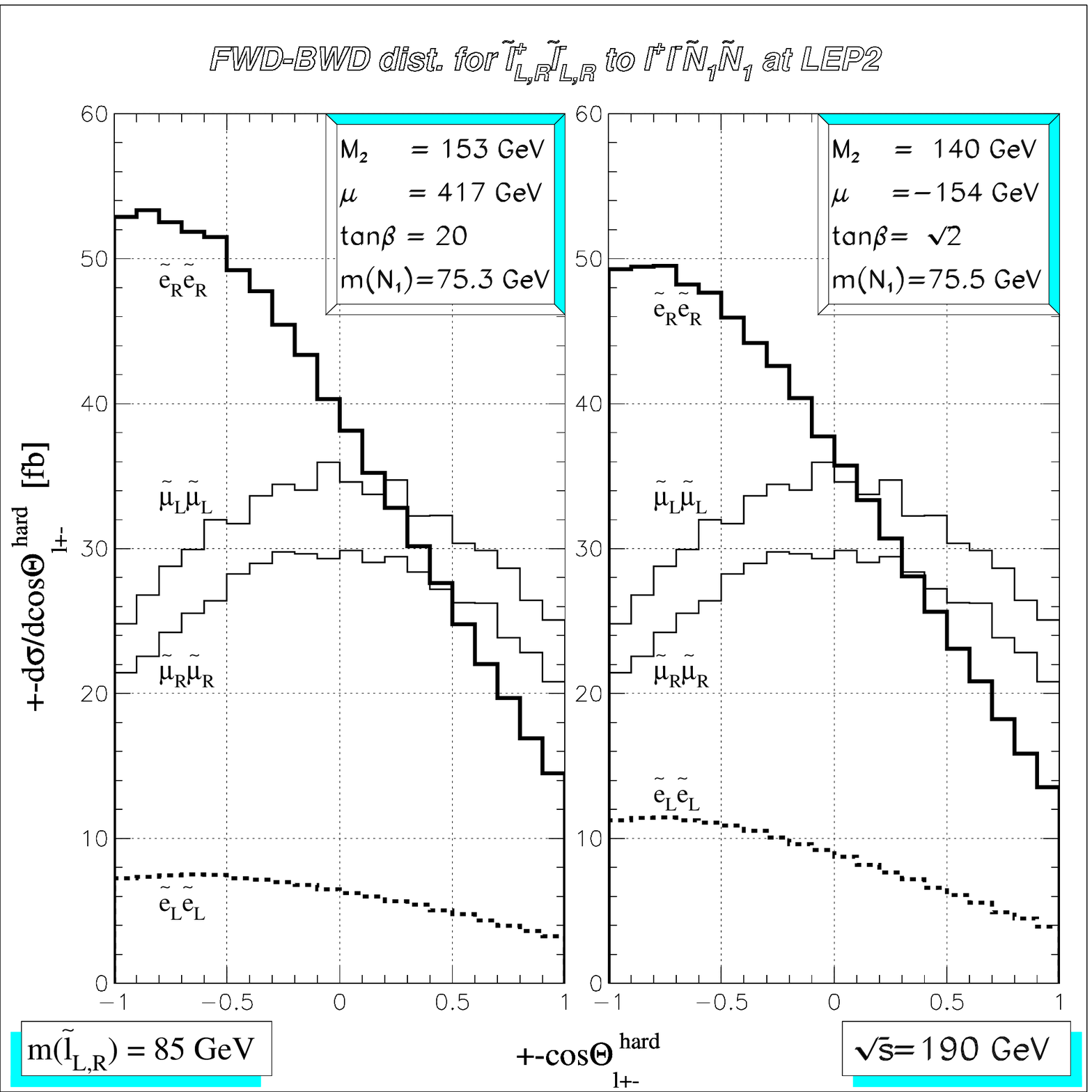}
}
\caption{The $\pm\cos\theta^{\rm hard}_{\ell^{\pm}}$ ($\ell = e$, $\mu$)
distribution [fb] for various slepton-pair production processes at LEP~II in
two examples of conventional GMSB models of Ref.~\protect\cite{AKM97}
(with $m_{\protect\sle{R}} = 85$~GeV and heavier $\protect\sle{L}$)
and two possible parameter choices for the HMSSM-I (with
$m_{\protect\sle{L}} = 85$~GeV and super-heavy $\protect\sle{R}$) where
$m_{\protect\n{1}} \simeq 75$~GeV.
The polar angle refers to the most energetic final lepton $\ell=e$ or
$\mu$ (after $\protect\sle{}\to\protect\n{1}\ell$ decay), with
respect to the $e^-$ beam direction. The $+$ $(-)$ sign is taken for the
cosine if such lepton is positively (negatively) charged.
The models considered are consistent with all collider limits (which
in particular force $m_{\protect\c{1}} > 125$~GeV), under the
hypothesis that $\protect\n{1} = $ NLSP and the
$\protect\n{1}\to\G\gamma$ decays occur inside the detector.
This plot has been obtained by explicitly generating 300K events
for each process with {\tt SUSYGEN 2.17} \protect\cite{susygen}.
Initial state radiation effects are included.
}
\label{sleptonfig}
\end{figure}

In Fig.~1, we show the $\pm\cos\theta^{\rm hard}_{\ell^{\pm}}$
($\ell = e$, $\mu$) distribution
for $\se{L}\se{L}$, $\smu{L}\smu{L}$
production in two possible parameter sets for the HMSSM-I with
$m_{\n{1}} \sim 75$~GeV, and $m_{\sle{L}} = 85$~GeV.
For comparison, analogous distributions are shown for the case of 
R slepton production, corresponding to two examples of conventional 
GMSB models with $\n{1} = $ NLSP amongst those of Ref.~\cite{AKM97}.
The two sets of parameters chosen are compatible with all collider 
limits and, in the GMSB case, with EWSB conditions as well%
\footnote{For the GSMB case, $\sle{R}$ can be light, but $\sle{L}$ pairs
          are always out of LEP~II reach. Only in a small number of cases
          can $\sepm{L}\semp{R}$ production be kinematically accessible.}.
Although the neutralino/chargino spectra and composition are similar, the
parameters otherwise have little in common. Hence, the two choices can
be thought as good examples of the general pattern%
\footnote{Note that in the GMSB models of Ref.~\protect\cite{AKM97}, if
          $\tan\beta\;\gtap$ 20, then the $\protect\sta{1}$ is always lighter
          than the $\protect\sle{R}$ by $\gtap$ 10~GeV. Thus,
          the model shown on the left side of Fig.~1 is a borderline
          $\protect\n{1} =$ (co-)NLSP case in conventional GMSB.}.
This polar angle parametrizes the F-B behavior of the charged leptons in
the final state. It is important that the decay $\n{1}\to\G\gamma$
occurs in the detector. The presence of hard, central (or displaced)
photons eliminates the SM physics background and also
allows the reconstruction of the $\n{1}$ and relevant selectron masses
for each event. For the parameters chosen in Fig.~1,
the final electrons (muons) have energies roughly in the range
5--15~GeV and can escape detection only when  $|\cos\theta|$ is close to 1.
For the smuons, the reactions can only proceed through $s$-channel
$\gamma$- or $Z$-exchange, and the distributions and total cross sections
are similar in the L and R cases. In contrast, for selectrons the
cross section and degree of F-B asymmetry is very
sensitive to the selectron handedness,  due to the fact that
$\se{L,R}\se{L,R}$ production also receives
contributions from $t$-channel neutralino exchange.
Amongst those, the light $\n{1}\sim\tilde{B}$
and the heavier $\n{2}\sim\tilde{W}^0_3$ dominate.
In the $\se{R}\se{R}$ case, only the former contributes, giving
rise to a 30--40\% increase in the total cross section,
when compared to smuon production.
In addition, there is a clear preference for producing a final
$e^-$ ($e^+$) in the same (opposite) hemisphere as the $e^-$ beam.
In contrast, the combined contributions to
$\se{L}\se{L}$ production not only produce destructive interference
and a total cross-section reduction, but also the resulting F-B
distribution is flatter. Still, in the two HMSSM-I cases of Fig.~1,
in a 500~pb$^{-1}$ run at $\sqrt{s} = 190$~GeV each LEP detector
should be able to observe 5--8 clean $e^+e^-\gamma\gamma + \miss{E}$ events.
One could then use the low cross section and smaller F-B asymmetry
to ascribe such events to $\sim$85~GeV $\se{L}$ pair production while 
inferring that $\se{R}$ pair production is above threshold, in contrast 
to most other SUSY models. Such a method should work similarly for
lighter selectrons, as long as the $\n{1}$ is the NLSP and heavier than 
70~GeV and the selectron is at least a few GeV heavier still%
\footnote{A similar disentangling technique would not be as effective
          in the case where selectrons, smuons and staus act as
          co-NLSPs and direct $\protect\se{}$, $\protect\smu{}$ decays
          to $\G$ are present, for two reasons. First, the lack of
          photons in the final state would make it harder to subtract the
          SM backgrounds, especially the one from $WW$. Unavoidably, severe
          cuts should be applied to the signal, with consequent loss of
          statistics. Second, the $\protect\tilde{B}-\protect\tilde{W}_3^0$
          pattern in the light neutralino sector would not
          hold in a model independent way.
          }.
For selectrons heavier than $\sim$ 85~GeV, lack of statistics will
hinder the above strategy for HMSSM pattern recognition.
However, as discussed below, $\gamma\gamma + \miss{E}$ events
from $\n{1}\n{1}$ production can provide another disentangling
tool. Further, sneutrino production followed
by $\tilde{\nu}\to\n{1}\nu$ must also occur in the HMSSM-I,
while with GMSB it need not. Finally, in both the HMSSM-I
and in GMSB models, one expects some
$\tau^+\tau^-\gamma\gamma + \miss{E}$ events from
$\sta{1}\sta{1}$ production, since typically
$m_{\n{1}} < m_{\sta{1}} < m_{\se{L,R}}$.

For the case of degenerate L and R selectrons at 85~GeV,
the two models considered in Fig.~1 would have a much higher
selectron production cross section of 650--800~fb
and a F-B asymmetry not too different from that of the
$\se{L}\se{L}$ case. However, such  degeneracy cannot be
realized in either the HMSSM or GMSB frameworks.

We now consider the HMSSM-I in the case where all slepton
production is above threshold at LEP~II,
while the $\n{1} $ is the  NLSP and is within reach. We assume the
$\n{1} $ decays into a photon and gravitino promptly%
\footnote{When the SSB scale is high enough so that the photon vertex
          is displaced, the near absence of SM background makes our
          arguments even stronger, while when the $\n{1} $ does not
          decay in the detector there is no visible signal.},
so that $e^+e^-\to\n{1}\n{1}$ with signature $\gamma\gamma + \miss{E}$
is the only observable SUSY signal. First of all, notice that the
production cross section is always too small to allow significant
occurrence of such events in the limited samples collected at LEP with c.m.
energies $\sqrt{s} = 161 - 172$~GeV, irrespective of the particular
model considered. This is in full agreement with the data on
$\gamma\gamma + \miss{E}$ events consistent with $\n{1}\to\G\gamma$
kinematics, with $m_{\n{1}}\; \gtap$ 70~GeV \cite{AKM97,AKM96,CDF-D0}.
However, the sensitivity of the $\n{1}\n{1}$ production process to
large L-R mass hierarchies in the selectron sector can still
distinguish the HMSSM-I from GMSB models in a
rather clean way,
provided that a good number of NLSP pairs are produced in the
forthcoming high-luminosity LEP~II run(s). As Fig.~2 shows,
if $m_{\n{1}} \ltap\ 85$~GeV, such production
can occur, but is not guaranteed.
\begin{figure}[h]
\centerline{
\epsfxsize=\textwidth
\epsffile{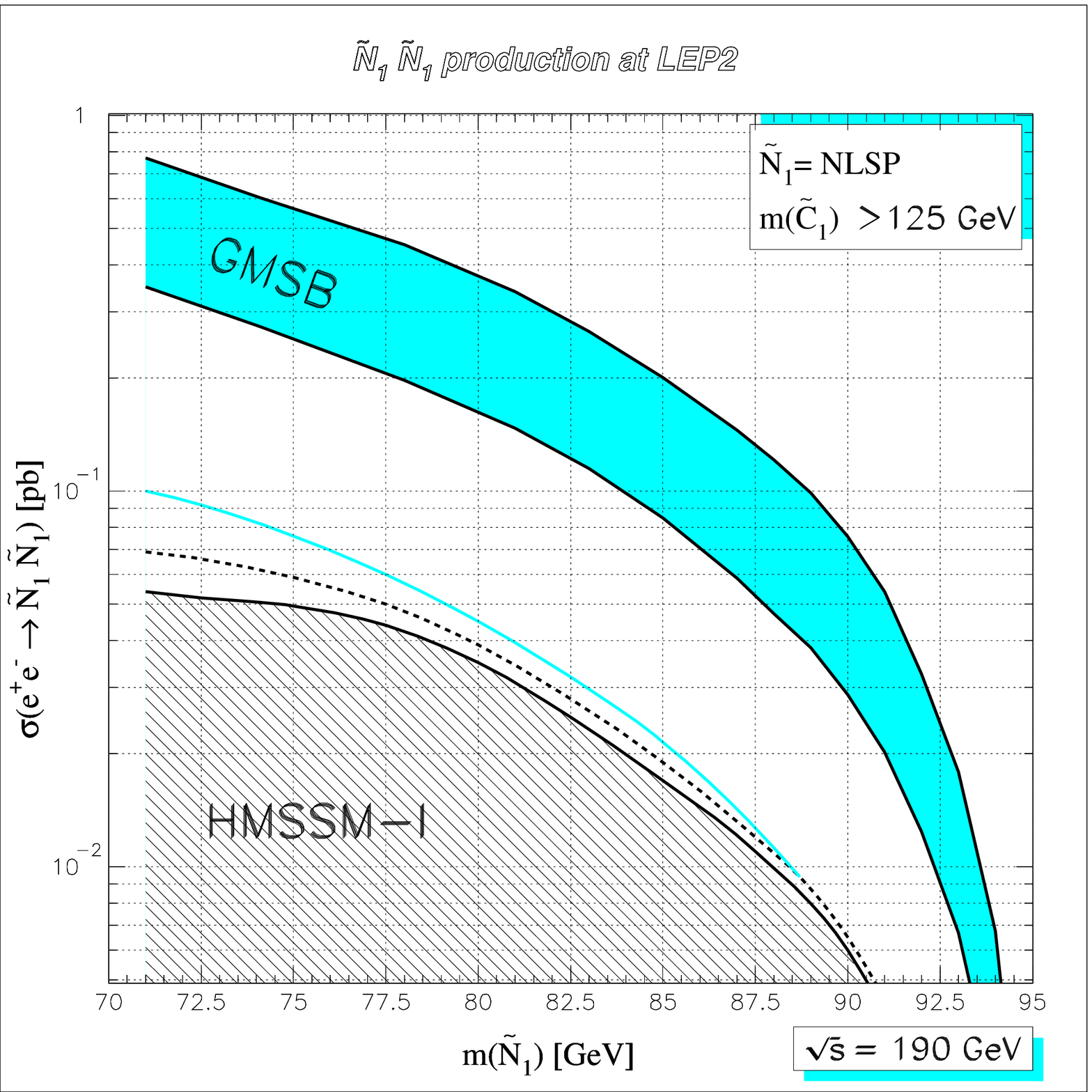}
}
\caption{Total cross section [pb] for $\protect\n{1}\protect\n{1}$
production at LEP~II in all the conventional GMSB models of
Ref.~\protect\cite{AKM97} (dark or blue region) and in HMSSM-I
models (hatched region) with $\protect\n{1} = $ NLSP and $\tan\beta > 1.2$.
All the models are consistent with all collider limits (which in particular
force $m_{\protect\c{1}} > 125$~GeV), under the hypothesis that 
the $\protect\n{1}\to\G\gamma$ decays occur inside the detector. 
The upper bound of the hatched region is given by the solid curve if 
$m_{\tilde{\nu}} > 95$~GeV $=[\protect\sqrt{s}/2]_{\rm LEP\;II}$;
by the dashed one if $m_{\protect\n{1}} < m_{\protect\tilde{\nu}} < 95$~GeV
$< m_{\protect\se{L}}$; by the grey (or blue) one if both sneutrino and L
selectron are below the threshold for pair production.
Initial state radiation effects are included.
}
\label{n1n1fig}
\end{figure}

Comparing the hatched region (HMSSM-I with no slepton
production at LEP~II) to the dark (or blue) one (all GMSB models of
Ref.~\cite{AKM97} with $\n{1}  =$ NLSP) it is evident that, for a given
$\n{1}$ mass, $\sigma(\n{1}\n{1})$ is always at least 4 to 6 times larger
in GMSB models. Indeed, after a suitable set of experimental cuts%
\footnote{Such cuts also avoid any possible contamination from events
          due to $\n{2}\n{2}$ production and subsequent double
          $\n{2}\to\n{1}\gamma$ radiative decay in SUGRA(-like) models.}
proposed in Refs.~\cite{AKM97,AKM96} to eliminate the SM
$\gamma\gamma\nu\bar{\nu}$ background, for \eg\ $m_{\n{1}} = 80$~GeV,
GMSB models with prompt $\n{1}\rightarrow \G\gamma$ decay predict more
than about 40 and up to 100 clean $\gamma\gamma + \miss{E}$ events,
while the HMSSM-I cannot give more than about 10.
The larger cross section in GMSB models is due to an effective upper bound on
$m_{\se{R}}$ for a given NLSP mass \cite{AKM97} and the large contribution to
$\n{1} $ pair production from $t$-channel $\se{R}$-exchange. (The
contribution from $\se{L}$-exchange is much smaller.)
Note that a clean measurement of the $\n{1}$ mass is possible
by using the kinematics of the production followed by
$\n{1}\to\gamma\G$ decay, after a selection of the events with the
highest photon energies.
However, as $m_{\n{1}}$ grows and the threshold for pair production
approaches, in the HMSSM-I  only a few $\gamma\gamma + \miss{E}$ events 
would be observed against a non-negligible background, so that such a 
discriminating method would be ambiguous. Also, one could be unlucky and
observe at most a very scarce signal, even for $m_{\n{1}} < 85$~GeV. 
On the other hand, detection of a relatively copious $\gamma\gamma + \miss{E}$
(or displaced-photon) signal, possibly for $m_{\n{1}}$ as large as
90~GeV, would exclude the HMSSM-I.
When $\sle{L}$ pair production is also allowed, the upper border of the
hatched region rises somewhat [\cfr\ grey (or blue) curve in Fig.~1] so that,
\eg\ for $m_{\n{1}} = 80$~GeV, one can have up to about 12 events
after cuts%
\footnote{In the models considered in Fig.~2 with $\tan\beta > 1.2$,
          this is only possible for $m_{\n{1}} \ltap $ 88.7~GeV,
          otherwise the sneutrinos become lighter than the
          $\n{1}$.}.
However, such a case is often realized when $m_{\sle{L}} \ltap$
85~GeV and one has other signals and disentangling tools available.
An intermediate, more involved case (\cfr\ dashed line in
Fig.~1) occurs when $m_{\n{1}} < m_{\tilde{\nu}} < 95$~GeV $< m_{\sle{L}}$.
Here, the presence of additional $\gamma\gamma + \miss{E}$
events from sneutrino production might give some background to
the $\n{1}\n{1}$ signal. However, it should be in principle
possible to distinguish sneutrino-originated events from
direct-neutralino production events, for instance by observing that
the former generally feature softer photons and a larger
missing energy. Also, when sneutrinos can be produced it is likely
that stau-pair production also occurs.
A final general observation is that photon
angular distributions are not a very good discriminant,
especially if the produced NLSPs are heavy,
since the final state pattern is dominated by
the kinematics of the $\n{1}\to\G\gamma$ decay which is isotropic
in the NLSP rest frame.

  A completely different scenario arises if the NLSP does not decay
inside the detector. Unlike the usual supergravity (SUGRA)
case however, the gravitino is the LSP, and cosmological
arguments do not require $\n{1}$ to be the NLSP. If, {\it e.g.} the
NLSP were $\tilde \tau_1$, SUSY events would not contain
missing energy but heavy charged tracks.
If the NLSP were $\n{1}$, then signatures of SUSY events would be somewhat
similar to those of SUGRA models, but even in this case the unusual
hierarchy of slepton masses could induce differences between the HMSSM-I
and conventional SUGRA models in the sparticle production cross sections
and in the branching fractions (BRs) for their decays. Unfortunately,
when the $\n{1} = $ NLSP is nearly stable, the limits on the
$\n{1}$ and $\c{1}$ masses are neither as general nor as stringent as in
the unstable case. Therefore, various compositions and spectra are
allowed for light $\n{1,2}$ and $\c{1}$, which makes it harder to find
a clean discriminant.

On the other hand, with a nearly stable $\n{1} =$ NLSP, heavier
neutralino-pair production, as well as $\cp{1}\cm{1}$ pair production,
could occur at LEP~II and provide a larger number of useful observables.
Indeed, in most HMSSM-I scenarios, we find a significant reduction of
the cross section for $\n{1}\n{2}$ production at LEP~II
with respect to conventional SUGRA models. This process
is especially sensitive to the right handed selectron
mass when the lighter neutralinos are mostly gauginos.
Also, perceivable differences in both the BRs and some final-state 
distributions are often present after the $\n{2}$ decay into the 
$\n{1}$. The angular distributions are especially affected when 
the decay process can be mediated by (on-shell) light L or R sleptons, 
as in the case $\n{2}\to\n{1}\ell^+\ell^-$. As an example, consider the
fact that a drastic enhancement of the $\n{2}$ visible leptonic BR can
be realized in a SUGRA model with $m_{\se{R}}\;\ltap\; m_{\n{2}}$, while
in the HMSSM-I this is much more difficult. Indeed, one has to force
$m_{\tilde{\nu}} < m_{\se{L}}\;\ltap\; m_{\n{2}}$, which often gives rise
to a substantial increase of the invisible fraction as well.
Although $\n{1}\n{2}$ searches allow exploration of a wider region of
the gaugino-higgsino parameter space than chargino searches do,
low cross sections and severe backgrounds could
render very difficult a post-discovery disentangling analysis
based only on neutralinos.

Signatures which use $\cp{1}\cm{1}$ production to distinguish the HMSSM-I
from SUGRA models are difficult to find, since $\cp{1}\cm{1}$ production
is insensitive to the R selectron mass. One might hope to  use, \eg,
a reduced hadronic chargino BR due to super-heavy L squarks as a signal
for the HMSSM-I, but squarks are quite heavy in many other models as well.
However, the HMSSM-II differs from SUGRA models in that
$\cp{1}\cm{1}$ production only proceeds through $s$-channel $\gamma$- or
$Z$-exchange, and, in contrast to  SUGRA models, the contributions
from $t$-channel $\tilde{\nu}$-exchange can never reduce the
cross-section at LEP~II.
Chargino BRs are also of interest, since a very large
leptonic fraction cannot be achieved in either version of the  HMSSM-II,
while such BRs could be large in either SUGRA or the HMSSM-I.
Particularly interesting is the case of the HMSSM-IIb, where
substantial lepton universality violation in chargino BRs and possibly 
in other quantities could be observed. If slepton production also occurs,
then the prospects for HMSSM/non-HMSSM disentangling should be brighter,
although SM-background reduction is still generally a more severe problem 
than when the decay $\n{1}\rightarrow \G\gamma$ is observed.

 As for HMSSM phenomenology at the Tevatron, the unusual HMSSM L-R mass
hierarchies are unlikely to generate striking signatures based on total
cross-sections or distributions of SUSY production processes, given our
hypothesis of relatively heavy squarks. However, some generic BR
arguments can still be made. For instance, relative to SUGRA models, the
HMSSM-I might give rise to an enhancement of the trilepton signal (when
the NLSP is a nearly stable neutralino) while the HMSSM-II (especially
HMSSM-IIa) would tend to give fewer trileptons.

One case where the HMSSM L-R slepton-mass hierarchy {\it is} important
at Tevatron occurs if the famous $e^+e^-\gamma\gamma + \miss{E_T}$ event
reported by CDF \cite{theevent} is a genuine SUSY discovery.
Assuming the NLSP is the  $\n{1}$ which decays promptly to $\G\gamma$,
the HMSSM-II (in particular HMSSM-IIa) cannot be compatible with the
event and other limits from $\gamma\gamma + X + \miss{E}$ inclusive
searches unless the event comes from $\se{R}\se{R}$ production with
$m_{\se{R}}\; \gtap$ 95~GeV.
However, the $\se{R}$ cannot be much heavier, since one
already expects less than a single $e^+e^-\gamma\gamma + \miss{E_T}$ event,
before experimental cuts, for $m_{\se{R}} = 95$~GeV \cite{eegg1,eegg2}.
However, in the HMSSM-I it is possible to interpret the event either as
$\se{L}\se{L}$ ($m_{\se{L}}\; \gtap$ 95~GeV) production or as
$\cp{1}\cm{1}$ production, followed by
$\c{1}\to\ell(\tilde{\nu}_{\ell}\to\nu_{\ell}\n{1})$ \cite{AKM97,AKM96,BKW}.
(The latter provides twice as many events with differently flavored
charged leptons.) Moreover, the $\n{1}$ mass is not correlated with the
selectron masses as in conventional GMSB models \cite{AKM97,AKM96,DTW1,BBCT}.
Thus, the number of expected additional non-standard events can be reduced.
In the chargino interpretation, models with
$m_{\c{1}} - m_{\tilde{\nu}} > 20$~GeV, which seems preferred by the
kinematics of the event, might be obtainable, whereas
$m_{\c{1}} - m_{\tilde{\nu}} > 20$~GeV is incompatible with $\n{1} = $
NLSP, \eg\ in the large class of GMSB models analyzed in Ref.~\cite{AKM97}%
\footnote{For a nearly-stable neutralino NLSP, an enhanced BR for the
          radiative $\n{2}\to\n{1}\gamma$ decay is required to generate
          the photons. Then similar arguments apply to the HMSSM and
          conventional SUGRA models. In particular, this renders
          problematic a chargino interpretation compatible with other
          limits \cite{eegg1,eegg2}.}.

Thus, for instance, observation of a larger number of
$\ell^+\ell^{(\prime) -}\gamma\gamma + \miss{E_T}$ events at the
Fermilab Main Injector, combined with at most a small signal at LEP~II from
$\n{1}\n{1}\to\gamma\gamma + \miss{E}$,  would be evidence for the HMSSM-I.
For this scenario, there would probably be additional
$\tau^+\tau^-\gamma\gamma$ events as well,
especially at Fermilab, coming from $\sta{1}$ production and decay.

In conclusion, we have considered general conditions under which
theories with multiple scales for the soft supersymmetry breaking
terms can avoid flavor changing neutral currents and large
Fayet-Iliopoulos terms, while maintaining natural electroweak symmetry
breaking. Such theories connect the two most mysterious aspects of
supersymmetric models, namely the physics of flavor and of
supersymmetry breaking.
We have  identified two new classes of models. These models have an
unusual hierarchy in scalar superpartner masses, with
either $m_{\se{L}}\ll m_{\se{R}}$,
$m_{\tilde{q}_L}\sim m_{\tilde{u}_R}\gg m_{\tilde{d}_R}$ (the HMSSM-I),  or
$m_{\se{L}}\gg m_{\se{R}}$, $m_{\tilde{q}_L}\sim m_{\tilde{u}_R}\ll
m_{\tilde{d}_R}$ (the HMSSM-II).
As an example, we have shown in some detail how the slepton mass pattern
$m_{\se{L}} \ll m_{\se{R}}$ can have distinctive
experimental consequences.\\

\begin{center}
{\bf Acknowledgements}
\end{center}
\noindent
We would like to thank Matt Strassler for collaboration at an early
stage of this work and useful discussions.
We are also grateful to Jonathan Feng, Gordy Kane, Graham Kribs and
Steve Martin for many useful discussions and suggestions.
A.N. was supported in part by the DOE under grant \#DE-FG03-96ER40956.
S.A. was supported mainly by the INFN, Italy.
S.A. also thanks the Particle Theory Groups at the University of
Michigan and at Fermilab for hospitality and additional support.


\end{document}